\documentclass[sigconf]{acmart}
\AtBeginDocument{%
  }

\copyrightyear{2026}
\acmYear{2026}
\setcopyright{cc}
\setcctype{by}
\acmConference[MM '26] {Proceedings of the 34th ACM International Conference on Multimedia}{November 10--14, 2026}{Rio de Janeiro, Brazil.}
\acmBooktitle{Proceedings of the 34th ACM International Conference on Multimedia (MM '26), November 10--14, 2026, Rio de Janeiro, Brazil}
\acmISBN{979-8-4007-2213-4/2026/11}
\acmDOI{10.1145/3767308.3834982}

\acmSubmissionID{mfp0487}

\usepackage{multirow}
\usepackage{graphicx}
\usepackage{balance}

\begin{document}

\title{Self-Reflective Multi-modal Reasoning for Short-Video Fake News Detection}


\author{Pinjie Xu}
 \authornote{Both authors contributed equally to this research. \ding{41}~Corresponding authors: Zhenxing Qian and Ce Li. }
\email{xupinjie321@outlook.com}
\affiliation{%
  \institution{China University of Mining and Technology - Beijing}
  \city{Beijing}
  \country{China}
}

\author{Yuzhou Yang}
 \authornotemark[1] 
\email{22110240074@m.fudan.edu.cn}
\affiliation{%
  \institution{Fudan University}
  \city{Shanghai}
  \country{China}
}
\orcid{0000-0001-6957-7682}

\author{Zhikai Tan}

\email{25213050086@m.fudan.edu.cn}
\affiliation{%
  \institution{Fudan University}
  \city{Shanghai}
  \country{China}
}
\orcid{0009-0001-5183-5695}

\author{Qichao Ying}
\email{shinydotcom@163.com}
\affiliation{%
  \institution{Fudan University}
  \city{Shanghai}
  \country{China}
}
\orcid{0000-0002-6527-2424}

\author{Zaiyang Yu}
\email{yuzaiyang@semi.ac.cn}
\affiliation{%
  \institution{The Institute of Semiconductors of the Chinese Academy of Sciences}
  \city{Beijing}
  \country{China}
}
\orcid{0000-0002-3425-1153}

\author{Ce Li}
\email{celi@cumtb.edu.cn}
\correspondingauthor
\affiliation{%
  \institution{China University of Mining and Technology - Beijing}
  \city{Beijing}
  \country{China}
}
\orcid{0000-0002-3081-6751}

\author{Zhenxing Qian}
\correspondingauthor
\email{zxqian@fudan.edu.cn}
\affiliation{%
  \institution{Fudan University}
  \city{Shanghai}
  \country{China}
}
\orcid{0000-0002-5224-6374}


\renewcommand{\shortauthors}{Pinjie Xu et al.}
\begin{abstract}
  Recent Fake News Detection (FND) pipelines increasingly leverage Large Language Models (LLMs) and Vision-Language Models (VLMs) for reasoning-based fake news detection.
  However, without ground-truth Chain-of-Thought (CoT) supervision, how to iteratively improve reasoning quality through self-reflection, 
  how such improved CoT can benefit downstream model finetuning, 
  and how to leverage large models to better connect single-sample fraudulent pattern discovery with cross-sample verification remain open problems.
  We propose \textbf{SRM-FND}, a self-reflective multi-modal reasoning framework for short-video fake news detection.
  SRM-FND constructs self-reflective and quality-improved reasoning through contrastive deliberation 
  with iterative root-cause diagnosis and corrective prompt refinement, where a Blind Analyst, 
  a Counter-Conclusion Reasoner, and a Self-Consistency Arbiter collaboratively retain discriminative rationales.
  It further adopts dual-phase topic-adaptive VLM finetuning for multi-modal grounding and lightweight topic specialization, and performs confidence-driven cross-sample review by retrieving co-event credible and suspicious witnesses for uncertain cases.
  Experiments on FakeSV and FakeTT demonstrate that SRM-FND outperforms strong baselines while yielding more reliable, interpretable predictions, 
  and also provide noticeable improvement in cross-dataset performance.
\end{abstract}

\ccsdesc[500]{Security and privacy~Social network security and privacy}

\keywords{Fake News Detection; Social Media; Multi-modal Large Models}


\maketitle

\begin{figure}
    \centering
    \includegraphics[width=\columnwidth]{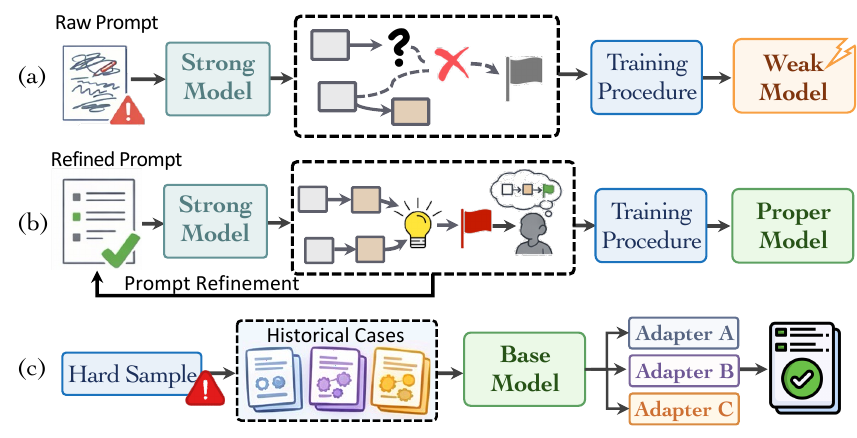}
    \caption{
    Paradigm comparison between existing works and the proposed SRM-FND.
    (a) Existing methods often construct CoT without sufficient mechanisms for quality assessment and correction.
    (b) SRM-FND constructs self-reflective and quality-improved reasoning through contrastive deliberation with iterative root-cause diagnosis and corrective prompt refinement.
    (c) Furthermore, SRM-FND introduces a cross-sample review mechanism to revisit samples that model predicts with less confidence by retrieving co-event training samples.
    Green items highlight the proposed mechanisms.
    }
    \label{fig:intro}
\end{figure}

\section{Introduction}
The proliferation of deceptive content on short-video platforms necessitates robust Fake News Detection (FND) systems, which identify suspicious content early 
and reduce its downstream impact~\cite{background2_systematicReview}.
Early studies primarily treat the task as a supervised classification problem, leveraging multimodal fusion of visual, textual, and metadata signals to predict veracity~\cite{jin2016novel,Leveraging_5,FND-Survey}.
While effective in capturing surface-level correlations, these approaches often lack explicit reasoning capability, making them vulnerable to spurious patterns and limiting their interpretability. 
Recent multimodal FND studies have increasingly leveraged large language models (LLMs) and vision-language models (VLMs) to generate
intermediate rationales or chain-of-thought (CoT) explanations for fake news verification~\cite{pan2023fact,li2025imrrf,wang2024explainable,wu2024sheep}. 
These methods improve transparency by exposing an explicit reasoning process and, in some cases, 
integrating structured retrieval or rationale supervision. However, current reasoning-based pipelines 
still leave two key questions insufficiently explored.
First, without ground-truth CoT labels, it remains unclear how reasoning quality
can be iteratively improved through self-reflection and how improved reasoning
can benefit downstream model fine-tuning.
Second, although modern VLMs enable cross-sample joint analysis, it remains
unclear when and how to connect the discovery of fraudulent patterns in
individual samples with cross-sample verification, especially for ambiguous
short-video cases that require event-level context.

In this paper, we propose SRM-FND, a self-reflective multimodal reasoning framework for short-video
fake news detection. 
SRM-FND operates as an iterative loop that alternates between reasoning
construction and model learning.
In each iteration, a contrastive deliberation process involving a Blind
Analyst, a Counter-Conclusion Reasoner, and a Self-Consistency Arbiter first
constructs self-reflective, higher-quality reasoning, while a root-cause
diagnosis and corrective prompt refinement mechanism progressively improves the
deliberation prompts across iterations.
The resulting CoT is then transferred into the detector through dual-phase
topic-adaptive VLM fine-tuning, which first aligns the backbone with the
multimodal structure of short-video news and then introduces topic-routed
lightweight adaptation for specialized discrimination.
At inference time, SRM-FND further introduces confidence-driven cross-sample
review, which automatically identifies uncertain predictions through
verdict-token confidence aggregation and revisits them by retrieving credible
and suspicious witnesses from related events.
We provide a paradigm comparison between existing methods and SRM-FND in
Figure~\ref{fig:intro}. 

We conduct extensive experiments on two popular short-video fake news
datasets, i.e., FakeSV~\cite{fakesv} and FakeTT~\cite{fakett_fakingrecipe}.
SRM-FND achieves 91.33 and 92.31 overall accuracy on the two datasets,
respectively, improving over the strongest prior baseline by 1.11 and 3.01
points.
Under cross-dataset transfer, SRM-FND achieves 76.20 and 72.24 overall
accuracy, outperforming the strongest prior baselines by 13.70 and 9.24
points, respectively.

In summary, the main contributions of this paper are as follows:
\begin{itemize}
    \item We introduce a self-reflective reasoning framework that performs contrastive CoT construction together with iterative root-cause diagnosis and corrective prompt refinement, enabling iterative improvement of reasoning quality without requiring ground-truth CoT labels.
    \item We propose a dual-phase topic-adaptive VLM fine-tuning scheme together with confidence-driven cross-sample review, which transfers improved reasoning into the detector and further connects single-sample analysis with event-level verification for uncertain cases.
\item Extensive experiments on FakeSV and FakeTT show that SRM-FND consistently outperforms strong baselines under both in-domain and cross-dataset settings, while providing more reliable and interpretable predictions.
\end{itemize}

\section{Related Works}
\label{sec:related_works}

\subsection{Multimodal Fake News Detection}
Existing multimodal FND schemes can be classified into two categories: content-oriented and evidence-based.
For content-oriented FND, the general paradigm is to transform news content into multimodal latent representations and directly learn veracity predictors without explicit evidence retrieval.
Early studies focus on unimodal discrepancies between real and fake news, e.g., visual artifacts in images~\cite{jin2016novel}, traces of manipulation~\cite{Leveraging_5}, and linguistic or stylistic patterns~\cite{FND-Survey}.
Subsequent multimodal approaches strengthen this line by jointly learning fused visual-textual representations~\cite{wei2022cross,WWW} or
exploiting pretrained vision-language models to assess cross-modal consistency~\cite{zhou2023multimodal,zhou2023multi,ying2023bootstrapping}.
For evidence-based FND, the general paradigm is to compare news with retrieved supporting materials and learn verification signals from news-evidence interactions.
Representative methods include early claim-evidence representation learning frameworks such as DeClarE~\cite{popat2018declare}, hierarchical attention architectures for fine-grained interaction modeling~\cite{ma2019sentence,wu2021evidence,vo2021hierarchical}, graph-based reasoning over distant semantic relations~\cite{xu2022evidence}, and iterative retrieval pipelines that refine evidence selection~\cite{liao2023muser,yang2024see}.
Evidence-based methods improve factual grounding, but they still often provide limited mechanisms to explicitly verify why the selected evidence should support the final decision, which can hinder interpretability, reliability, and robustness under domain shift.

\begin{figure*}
    \centering
    \includegraphics[width=\linewidth]{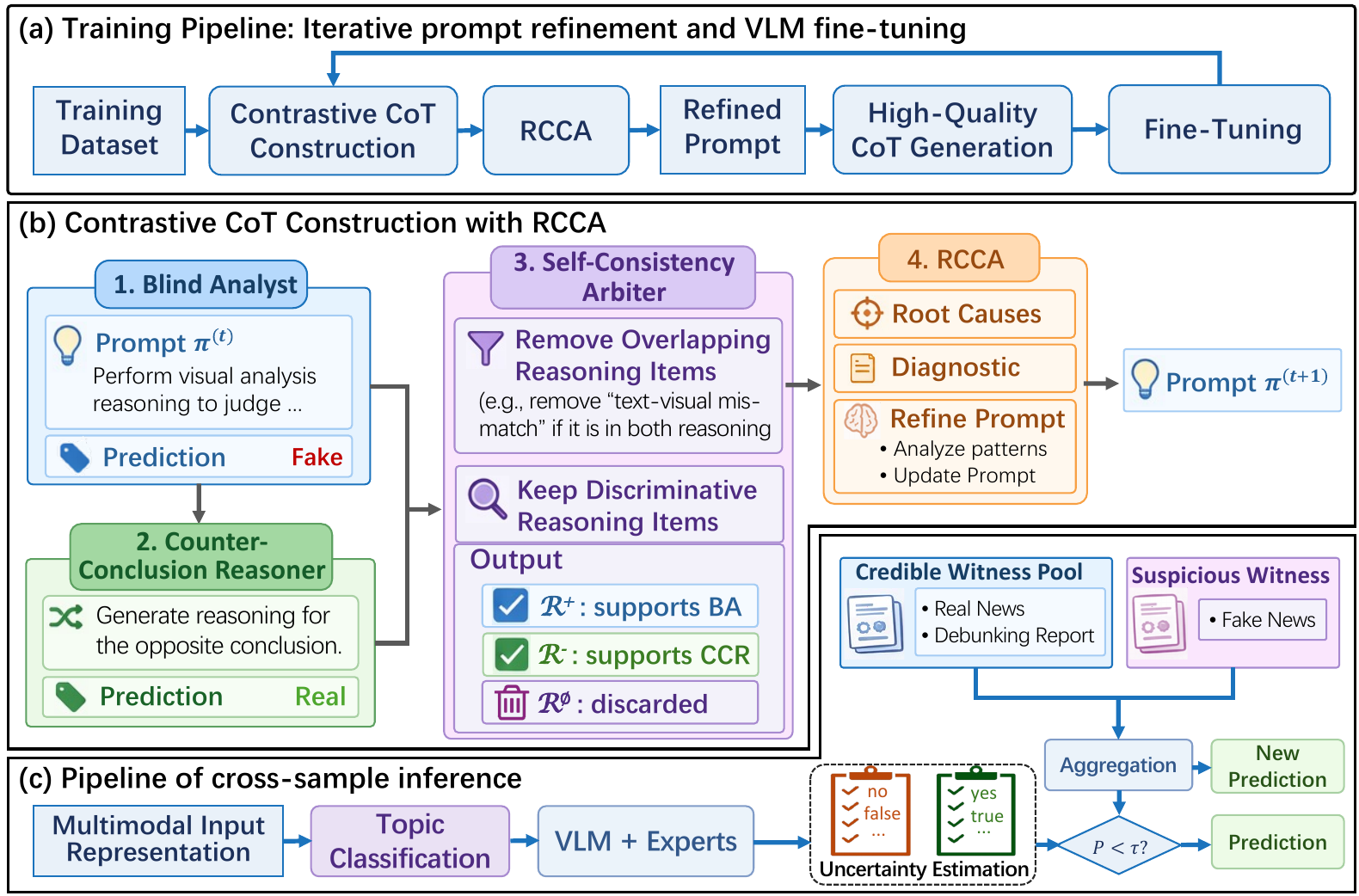}
    \caption{
    The SRM-FND framework consists of three interleaved components.
    First, the Blind Analyst and Counter-Conclusion Reasoner generate opposing reasoning chains, while the Self-Consistency Arbiter removes overlapping arguments and retains discriminative reasoning.
    RCCA then diagnoses pipeline failures and iteratively updates only the Blind Analyst's prompt.
    Second, SRM-FND adapts a backbone VLM through multimodal alignment pre-training and topic-routed expert fine-tuning for topic-specialized multimodal fake news detection.
    Third, SRM-FND performs confidence-driven cross-sample review for uncertain cases by retrieving co-event samples for final adjudication.
    }
    \label{fig:pipeline}
\end{figure*}
\subsection{Reasoning-augmented Fake News Detection}

Recent advances introduce reasoning-augmented paradigms that leverage large language models (LLMs) to generate intermediate rationales for fact verification. Instead of directly mapping inputs to labels, these methods incorporate explicit reasoning processes to improve interpretability and decision quality.
A representative direction constructs structured reasoning procedures. For example, program-guided frameworks decompose verification into executable reasoning steps generated by LLMs, enabling more controllable inference~\cite{pan2023fact}. 
Some methods further combine rationale generation with supervised learning. For instance, methods may first use LLMs to produce explanations or competing evidence sets and then train classifiers on both original inputs and generated rationales to improve prediction reliability~\cite{wang2024explainable,li2025imrrf}. Additionally, data augmentation strategies based on LLM-generated rewrites have been explored to reduce stylistic bias and encourage models to focus on factual content~\cite{wu2024sheep}.
While reasoning-augmented methods provide improved interpretability, they still face several challenges. 
The generated rationales can be noisy or hallucinated, potentially introducing instability into the decision process. Furthermore, many existing approaches rely on single-pass reasoning without mechanisms for self-correction or experience accumulation, limiting their robustness when handling ambiguous or previously unseen cases.
This motivates us to explore self-reflective reasoning for FND.

\section{Proposed Method}
\label{sec:method}

\noindent\textbf{Approach Overview.}
Fig.~\ref{fig:pipeline} depicts the pipeline of the proposed SRM-FND approach.
The framework is composed of three interleaved components:
(i)~\textit{Contrastive CoT Construction with RCCA},
which synthesizes self-reflective, higher-quality reasoning for every training instance through
a multi-perspective deliberation pipeline and iteratively refines the underlying reasoning
prompts to address systematic generation failures;
(ii)~\textit{Dual-Phase Topic-Adaptive VLM Fine-Tuning}, which first
grounds the base VLM in the heterogeneous multimodal
structure of short-video news through multimodal alignment pre-training and then
attaches topic-routed expert branches for fine-grained discrimination;
and (iii)~\textit{Confidence-Driven Cross-Sample Review}, which identifies ``borderline'' predictions
through dual-token-set probability thresholding
and adjudicates them by retrieving co-event credible and suspicious witnesses from related events.

\subsection{Multi-modal Input Preprocessing and Topic-Oriented Data Organization}

A multimodal short-video news sample is denoted by
$\mathbf{s} = (V, T, M) \in \mathcal{D}$, where $V$ represents the video content
(sampled keyframes), $T$ denotes the associated textual content (the title,
automatic speech recognition transcript, and on-screen text), and $M$ encodes
auxiliary metadata (publication time, engagement statistics, commenter content,
and publisher profile).
Each sample $\mathbf{s}$ carries a binary ground-truth label $y \in \{0,1\}$
and a topic label $k \in \{1,2,3,4\}$ corresponding to \textit{Social Safety, Hard News,
Specialized Knowledge, and Life \& Trivia}, respectively. Details are as follows.

\noindent\textbf{Multimodal Input Preprocessing.}
To support different types of multimodal fake news samples, we first build a
uniform preprocessing pipeline that converts raw video news into the structured
representation described above.
For speech content, each video is transcribed via automatic speech recognition
(ASR) and filtered for transcription quality before being incorporated into $T$.
For on-screen text, Optical Character Recognition (OCR) annotations are further refined through an LLM-guided
denoising procedure to remove watermarks and garbled characters that would
otherwise introduce noise into the textual input.
For metadata, publisher information and engagement signals are extracted from
platform-provided fields to form the metadata component $M$.

\noindent\textbf{Topic-Oriented Data Organization.}
Fake news videos exhibit pronounced topic heterogeneity,
and prior research in FND~\cite{nan2021mdfend,zhu2022memory} confirms that topical diversity has a material impact on how misinformation is constructed
and how it should be identified.
The four topic categories adopted in this work are grounded in well-established
journalism taxonomies~\cite{reinemann2012hard,baum2007soft,zhou2026finefake}.
For example, 
Social Safety and Hard News correspond to the hard-news spectrum where verifiability
is high and fabrications tend to be forgery- or provenance-based~\cite{reinemann2012hard}, 
whereas
Specialized Knowledge and Life \& Trivia occupy the soft-news end where
misinformation more commonly exploits low audience expertise or emotional salience~\cite{baum2007soft}.
A coarser grouping would conflate meaningfully distinct deception patterns, while
a finer-grained partition would excessively fragment the training data available to each expert
and render per-topic adaptation unreliable.
The four-category scheme empirically strikes a practical balance between discriminative
granularity and data sufficiency for per-topic model training.

For each sample, a topic is assigned via VLM classification, and we activate
only the model corresponding to the selected topic.
Because ground-truth topic labels are unavailable, topic-classification accuracy
cannot be measured. We therefore use the same VLM for topic classification during
framework preparation and inference, ensuring that any systematic classification
bias is consistent across all stages.

\subsection{Contrastive CoT Construction with Root Cause and Corrective Action (RCCA)}
\label{sec:cot}

Constructing high-quality CoT for a VLM requires more than a single-pass
prompting strategy.
One common issue is ambiguity (or hallucination), in which the same rationale can even be
used to support both real and fake predictions, and incorporating such
ambiguous signals can reduce the discriminativeness of the learned
representation.
A further challenge is how to adaptively update the prompt via a self-reflective mechanism.
We address these challenges through contrastive CoT construction with RCCA.

\noindent\textbf{CoT Construction with Contrastive Deliberation.}
\label{sec:trirole}
The deliberation pipeline instantiates three LLM roles.
The \textit{Blind Analyst}, or $\mathcal{B}$, receives the full multimodal input of $\mathbf{s}$
and produces a topic-specific reasoning chain together with a binary verdict
$c_{\mathrm{BA}} \in \{\textsc{Real}, \textsc{Fake}\}$, while the ground-truth
label $y$ is withheld.
Next, the \textit{Counter-Conclusion Reasoner}, or $\mathcal{R}$, is given the opposite
conclusion $\neg c_{\mathrm{BA}}$ as posterior knowledge and generates an alternative
reasoning chain supporting that conclusion; it shares the same prompt with $\mathcal{B}$.
Both roles are required to provide a list of perspectives supporting their respective conclusions.
Afterward,
the \textit{Self-Consistency Arbiter}, or $\mathcal{A}$, compares both reasoning chains, removes
semantically overlapping arguments, and outputs ${R}^{+}$,
${R}^{-}$, and the eliminated set ${R}^{\emptyset}$.
Using the ground-truth label $y$, we retain ${R}^{+}$ if $y$ is
consistent with $c_{\mathrm{BA}}$, and otherwise retain ${R}^{-}$.
Samples with an empty correct-side reasoning list are retried up to
$T_{\text{retry}}$ times and then excluded.
Perspectives are considered ambiguous if they appear on both sides with semantically equivalent meanings.
At each iteration $t$, every training sample with a non-empty selected CoT
constitutes the dataset $\mathcal{C}^{(t)}$, where each record stores
$\mathbf{s}$, its retained reasoning chain, and the aligned label $y$.
For datasets containing debunking videos, we incorporate these samples into the corresponding topic expert's training corpus and map their verdict labels to \textit{Real}. Their CoTs are generated by the VLM using the debunking content as posterior evidence.

\noindent\textbf{Root Cause and Corrective Action (RCCA).}
\label{sec:rcca}
The prompt $\pi^{(t)}$ of the Blind Analyst $\mathcal{B}$ and the resulting failure patterns on the
training set define a closed-loop optimization procedure.
Each RCCA cycle proceeds as follows.
The current prompt $\pi^{(t)}$ drives the full deliberation pipeline over all
training samples, and each sample's outcome is compressed into a single sentence that
summarizes the CoT construction process.
To ensure broad diagnostic coverage, failure cases are first stratified by
event label so that diverse events are represented, then randomly sampled to
yield up to $N_{\text{diag}}$ summaries, bounded by the context window.
These summaries, together with $\pi^{(t)}$, form a structured diagnostic
document from which an optimizer LLM identifies root causes and proposes a
refined prompt $\pi^{(t+1)}$.
To prevent overfitting to the sampled failures, the optimizer prompt explicitly
instructs the model to treat identified failure patterns as a supplementary
reference rather than the sole basis for revision.
Only $\mathcal{B}$'s prompt is updated per cycle. The Counter-Conclusion
Reasoner $\mathcal{R}$ and Self-Consistency Arbiter $\mathcal{A}$ are held fixed to provide single-variable
attribution for observed improvements.
Topics are optimized independently, with cycle count determined by convergence
of per-topic validation accuracy.
Note that RCCA-driven CoT construction and refinement are performed solely on the
training split and separately for each topic.

\subsection{Dual-Phase Topic-Adaptive VLM Fine-Tuning}
\label{sec:lora}

Using the generated CoT, we adopt dual-phase topic-adaptive VLM fine-tuning.
We first fine-tune the backbone on the multimodal structure of short-video news
and then perform lightweight, topic-routed fine-tuning for specialized discrimination.
Low-rank adaptation (LoRA)~\cite{hu2022lora} is used in this stage.

\noindent\textbf{Multi-modal Alignment Pre-Training.}
The first tuning phase adapts the VLM to the domain-specific multi-modal input
structure through a reconstruction objective entirely decoupled from fake news labels.
Given a sample $\mathbf{s} = (V, T_{\neg \text{title}}, M)$, where the news
title and core event keywords are withheld from the input, the model is trained
to predict these held-out fields autoregressively.
The event keywords are sourced from dataset-provided annotations when available.
For datasets lacking such labels, they are extracted via LLM inference as a
fallback.
This compels the model to establish meaningful correspondences across video frames,
ASR transcripts, on-screen text, engagement metadata, and publisher signals
before any discriminative objective is introduced.
The pre-training loss follows the standard causal language modeling criterion:
\begin{equation}
\mathcal{L}_{\text{align}} = -\sum_{i} \log P_{\theta}(w_i \mid w_{<i}, V, T_{\neg \text{title}}, M),
\end{equation}
where $\{w_i\}$ are tokens in the withheld title and keywords.

\noindent\textbf{Topic-Routed Expert Fine-Tuning.}
\label{sec:finetune}
Next, based on the pre-trained backbone, we attach topic-specific lightweight
parameter branches $\{\Delta\theta_k\}_{k=1}^{K}$ to the VLM, one
per topic category, and train them concurrently on the CoT dataset
$\mathcal{C}^{(t)}$ from the current RCCA iteration.
Each sample's supervision consists of a conclusion-first structured response.
The model first emits the binary verdict token, then generates the supporting
CoT chain.
This ordering ensures that the judgment conditions the subsequent reasoning generation.
The training objective combines two cross-entropy losses: one for the verdict token
and one for the CoT sequence:
\begin{equation}
\mathcal{L} = \mathcal{L}_{\text{cls}} + \alpha \mathcal{L}_{\text{CoT}},
\end{equation}
where $\alpha$ balances the CoT supervision term.
At each RCCA iteration $t$, the CoT dataset $\mathcal{C}^{(t)}$ is used to
re-instantiate this fine-tuning phase from the pre-trained backbone, producing
an updated detection model whose improved CoT is reflected in
both verdict accuracy and reasoning quality.

\subsection{Confidence-Driven Cross-Sample Review}
\label{sec:rag}

This detector mainly investigates each sample individually.
Although such single-sample analysis can already reveal many fraudulent signals,
some cases still require explicit cross-sample inspection for further
verification.
Directly feeding excessive cross-sample information into a VLM, however, may
make it harder for the model to efficiently identify useful clues.
This motivates us to first identify samples that require further cross-sample
analysis and then guide the VLM to perform targeted cross-checking within a
narrowed scope.

\noindent\textbf{Inference Uncertainty Estimation.}
\label{sec:unc}
For each test sample $\mathbf{s}$, the fine-tuned model produces a verdict token
distribution over its vocabulary.
Let $\mathcal{V}^{+}$ denote the predefined set of tokens semantically aligned
with a \textit{real} verdict, e.g., \textit{true}, \textit{real},
\textit{correct}, and let $\mathcal{V}^{-}$ denote the set aligned with a
\textit{fake} verdict, e.g., \textit{false}, \textit{fake}, \textit{incorrect}.
Both sets are manually curated with respect to the VLM's tokenizer. For
multi-token words, the joint probability of the constituent subword tokens is
used. The full token lists are provided in the supplementary material.
The aggregated confidence scores are defined as $P^{+} = \sum_{v \in \mathcal{V}^{+}} p(v \mid \mathbf{s})$
and $P^{-} = \sum_{v \in \mathcal{V}^{-}} p(v \mid \mathbf{s})$.
A prediction is flagged as uncertain if neither aggregated score reaches
the predefined threshold $\tau$, i.e.,
$\text{uncertain}(\mathbf{s}) \iff P^{+} < \tau \;\wedge\; P^{-} < \tau$.
We observe empirically that the fine-tuned model assigns high confidence to
the vast majority of test samples, making this criterion selective rather than
indiscriminate. The threshold $\tau$ is set empirically based on
validation-set performance.

\noindent\textbf{Cross-Sample Review.}
For samples flagged as uncertain, SRM-FND retrieves co-event witnesses
from the training corpus.
Each sample is associated with an event label.
When event annotations are available, they are used directly.
Otherwise, concise event descriptions generated by the VLM serve as fallback
event labels.
Samples sharing the same or a closely related event label with the uncertain
query and carrying high-confidence predictions, i.e., $\max(P^{+}, P^{-}) \geq \tau$,
are partitioned into two complementary witness pools.
The pool of \textit{credible witnesses} comprises high-confidence real news and
authoritative debunking reports from the same event cluster, and the
pool of \textit{suspicious witnesses} comprises high-confidence fake news samples
from the same cluster.
If no co-event witnesses are retrieved, cross-sample review is bypassed and the
prediction associated with the higher aggregated score is directly adopted as the
final verdict.
Otherwise, adjudication proceeds in two complementary steps.
For the credible witness pool, the model is prompted to first synthesize a
factual account of the event by drawing on the corroborated real news and
debunking sources, and then cross-examine the uncertain sample against this
established account to assess factual consistency.
For the suspicious witness pool, the model is prompted to compare the uncertain
sample against each fake news witness individually, identifying shared fabrication
patterns, narrative deformations, or visual forgery traces that would link the
query to the same deceptive event cluster.
The final cross-sample verdict is produced by a single call to the adjudication
VLM, which integrates both analyses.

\begin{table}[t]
  \centering
    \caption{Topic-wise statistics of FakeSV and FakeTT. Samples are categorized through VLM-based topic classification.}
\setlength{\tabcolsep}{1mm}
  \begin{tabular}{lrrrr|rrrr}
    \toprule[1.5pt]
    \multirow{2}{*}{\textbf{Topic}} & \multicolumn{4}{c}{\textbf{FakeSV}} & \multicolumn{4}{c}{\textbf{FakeTT}} \\
    \cline{2-9}
    & Train & Val. & Test & Deb. & Train & Val. & Test & Deb. \\
    \hline
    \textit{Life \& Trivia} & 525 & 121 & 159 & 392 & 561 & 105 & 75 & - \\
    \textit{Social Safety} & 1,470 & 296 & 223 & 715 & 146 & 35 & 26 & - \\
    \textit{Hard News} & 285 & 52 & 56 & 256 & 361 & 80 & 142 & - \\
    \textit{Specialized} & 256 & 77 & 104 & 508 & 325 & 79 & 56 & - \\
    \hline
    \textbf{Total} & \textbf{2,536} & \textbf{546} & \textbf{542} & \textbf{1,871} & \textbf{1,393} & \textbf{299} & \textbf{299} & - \\
    \bottomrule[1.5pt]
  \end{tabular}
  \label{tab:data_stats}
\end{table}

\begin{table*}[t]
  \centering
  \caption{In-domain and cross-dataset performance on FakeSV and FakeTT. Best results are shown in boldface, and second-best results are underlined. $\downarrow$ denotes the ACC drop. Cross-dataset baseline results are taken from published reports or reproduced using official open-source repositories. The CoT corpus is generated by analyst roles instantiated with GPT-5.1.}
    \setlength{\tabcolsep}{1.2mm}
  \begin{tabular}{l|rrr|rrrr|rrr|rrrr}
    \toprule[1.5pt]
    \textbf{Dataset} & \multicolumn{3}{c}{\textbf{FakeSV}$\rightarrow$\textbf{FakeSV}} & \multicolumn{4}{c}{\textbf{FakeTT}$\rightarrow$\textbf{FakeSV}} & \multicolumn{3}{c}{\textbf{FakeTT}$\rightarrow$\textbf{FakeTT}} & \multicolumn{4}{c}{\textbf{FakeSV}$\rightarrow$\textbf{FakeTT}} \\
    \hline
    \textbf{Model} & ACC & M-F1 & M-R & ACC & $\downarrow$ & M-F1 & M-R & ACC & M-F1 & M-R & ACC & $\downarrow$ & M-F1 & M-R \\
    \hline
    Qwen3-VL~\cite{bai2025qwen3} & 64.21 & 60.79 & 61.52 & - & - & - & - & 65.96 & 63.14 & 64.06 & - & - & - \\
    InternVL3.5~\cite{chen2024internvl} & 56.27 & 37.87 & 50.39 & - & - & - & - & 55.18 & 55.10 & 63.95 & - & - & - \\
    GPT-5.1 & 65.87 & 58.91 & 61.77 & - & - & - & - & 70.23 & 69.09 & 71.63 & - & - & - & - \\
    Claude-4.5 & 66.05 & 59.36 & 62.03 & - & - & - & - & 68.56 & 68.43 & 74.97 & - & - & - & - \\

    \hline
    FANVM~\cite{FANVM} & 79.88 & 78.91 & 78.42 & 51.43 & 28.45  & 42.21 & 51.39  & 71.91 & 70.85 & 73.90 & 57.48 & 14.43 & 45.13 & 50.65 \\
    SV-FEND~\cite{fakesv} & 80.81 & 80.19 & 79.84 & 53.53 & 27.28 & 50.36 & 53.56  & 77.26 & 75.55 & 77.13 & 63.00 & 14.26 & 60.06 & 60.07 \\
    FakingRecipe~\cite{fakett_fakingrecipe} & 84.69 & 84.39 & 84.25 & 62.50 & 22.19 & 62.24 & 62.49  & 79.26 & 77.53 & 78.89 & 61.09 & 18.17 & 55.74 & 60.45 \\
    ExMRD~\cite{ExMRD} & 86.90 & 86.52 & 86.13 & 50.52 & 36.38 & 34.77 & 50.47 & 84.28 & 83.13 & 85.19 & 60.04 & 24.24 & 54.52 & 53.69 \\
    FakeSV-VLM~\cite{wang2025fakesv} & \underline{90.22} & \underline{89.97} & {89.64} & - & - & - & -  & {89.30} & {87.98} & {88.17} & - & - & - & -  \\
    \hline
    \textbf{SRM-FND (Qwen3-VL)} & \textbf{91.33} & \textbf{91.05} & \textbf{90.54} & \textbf{76.20} & 15.13 & \textbf{76.20} & \textbf{77.28}  & \textbf{92.31} & \textbf{91.25} & \textbf{90.93} & \textbf{72.24} & 20.07 & \textbf{71.51} & \textbf{75.42} \\
    \textbf{SRM-FND (InternVL3.5)} & {90.04} & {89.84} & \underline{89.66} & \underline{64.94} & 25.10 & \underline{63.50} & \underline{68.39}  & \underline{91.30} & \underline{89.97} & \underline{89.16} & \underline{70.23} & 21.07 & \underline{69.63} & \underline{73.92} \\
    \bottomrule[1.5pt]
  \end{tabular}
  \label{tab:main_results}
\end{table*}

\begin{table*}[t]
  \centering
  \caption{Per-topic performance of SRM-FND on FakeSV and FakeTT. The upper block corresponds to SRM-FND (Qwen3-VL), and the lower block corresponds to SRM-FND (InternVL3.5). $\downarrow$ denotes the ACC drop.}
  \begin{tabular}{l|l|rrr|rrrr|rrr|rrrr}
    \toprule[1.5pt]
    \multirow{2}{*}{\textbf{Model}} & \multirow{2}{*}{\textbf{Topic}}
    & \multicolumn{3}{c}{\textbf{FakeSV}$\rightarrow$\textbf{FakeSV}}
    & \multicolumn{4}{c}{\textbf{FakeTT}$\rightarrow$\textbf{FakeSV}}
    & \multicolumn{3}{c}{\textbf{FakeTT}$\rightarrow$\textbf{FakeTT}}
    & \multicolumn{4}{c}{\textbf{FakeSV}$\rightarrow$\textbf{FakeTT}} \\
    \cline{3-16}
    &
    & ACC & M-F1 & M-R 
    & ACC & $\downarrow$ & M-F1 & M-R 
    & ACC & M-F1 & M-R 
    & ACC & $\downarrow$ & M-F1 & M-R \\
    \hline

    \multirow{4}{*}{\rotatebox{90}{\textbf{Qwen3-VL}}}
    & \textit{Life \& Trivia} & 88.68 & 83.51 & 80.44 & 74.21 & 14.47 & 65.47 & 65.12 & 93.33 & 91.34 & 89.55 & 80.00 & 13.33 & 70.86 & 68.65 \\
    & \textit{Social Safety} & 92.38 & 90.81 & 93.73 & 82.06 & 10.32 & 72.16 & 69.36 & 92.31 & 48.00 & 50.00 & 53.85 & 38.46 & 41.35 & 52.08 \\
    & \textit{Hard News} & 92.86 & 92.22 & 92.22 & 83.93 & 8.93 & 83.01 & 84.17 & 91.55 & 78.87 & 76.16 & 80.99 & 10.56 & 63.00 & 65.37 \\
    & \textit{Specialized} & 92.31 & 79.67 & 75.82 & 62.50 & 29.81 & 54.77 & 71.98 & 92.86 & 92.77 & 92.77 & 48.21 & 44.65 & 39.74 & 52.84 \\
    
    \hline

    \multirow{4}{*}{\rotatebox{90}{\textbf{InternVL3.5}}}
    & \textit{Life \& Trivia} & 86.16 & 80.25 & 77.95 & 46.54 & 39.62 & 46.53 & 61.60  & 89.33 & 87.46 & 89.68 & 76.00 & 13.33 & 67.11 & 65.87  \\
    & \textit{Social Safety} & 91.93 & 89.39 & 88.54 & 79.82 & 12.11 & 63.94 & 62.41  & 96.15 & 88.94 & 97.92 & 73.08 & 23.07 & 52.97 & 62.50 \\
    & \textit{Hard News} & 91.07 & 90.56 & 91.94 & 66.07 & 25.00 & 65.97 & 73.61 & 92.25 & 78.88 & 74.19 & 72.54 & 19.71 & 58.28 & 65.28 \\
    & \textit{Specialized} & 91.35 & 73.93 & 68.68 & 60.58 & 30.77 & 51.50 & 64.29 & 89.29 & 89.23 & 89.55 & 55.36 & 33.93 & 49.55 & 59.68 \\

    \bottomrule[1.5pt]
  \end{tabular}
  \label{tab:topic_results}
\end{table*}

\section{Experiments}
\label{sec:experiments}

\subsection{Implementation and Experimental Settings}
\label{sec:implementation_detail}
\noindent\textbf{Framework Settings.}
The backbone VLM can be selected from popular open-source models such as
Qwen3-VL-8B-Thinking~\cite{bai2025qwen3} or
InternVL3.5-8B~\cite{chen2024internvl}.
All analyst roles and the RCCA optimizer can be instantiated with more powerful
models such as GPT-5.1 or Claude-4.5.
All deliberation roles and the RCCA optimizer share the same underlying model to
ensure stylistic and capability consistency across pipeline stages.
Speech content is transcribed with Qwen3-ASR~\cite{shi2026qwen3} before being
incorporated into the textual input.
For each video, we evenly sample at most 10 keyframes from the full video as
visual input to the deliberation pipeline.
The retry cap for empty correct-side reasoning lists is
$T_{\text{retry}}\!=\!3$, and the maximum number of diagnostic failure summaries
per RCCA cycle is $N_{\text{diag}}\!=\!100$.
In the multimodal alignment pre-training phase, a shared LoRA with rank 64
and LoRA scaling factor 128 is applied uniformly across all transformer layers,
while in the topic-routed fine-tuning phase each of the $K\!=\!4$ topic experts
uses a LoRA with rank 64
applied to all linear layers except the ViT encoder and multimodal aligner.
Pre-training and fine-tuning use the AdamW~\cite{adamw} optimizer with a learning
rate of $5\times10^{-5}$, a weight decay of 0.1, and the hyperparameter
$\alpha\!=\!0.5$, and continue until convergence.
After each RCCA iteration, fine-tuning is re-instantiated from the pre-trained
backbone and trained on the training split together with the CoT corpus
$\mathcal{C}^{(t)}$, while the validation split is used only for monitoring fine-tuning convergence and model selection.
The uncertainty threshold is set to $\tau\!=\!0.6$.
Model training is conducted on eight NVIDIA A100 GPUs.

We run a small number of RCCA iterations and select the final
checkpoint according to validation-set accuracy.
This design keeps CoT construction and refinement restricted to the training
split, while the validation split is reserved for checkpoint comparison and
hyperparameter selection rather than reasoning generation.
The test set is never used during the framework-preparation stage.
This helps balance performance and overall inference cost.


\begin{figure*}[!t]
  \includegraphics[width=\linewidth]{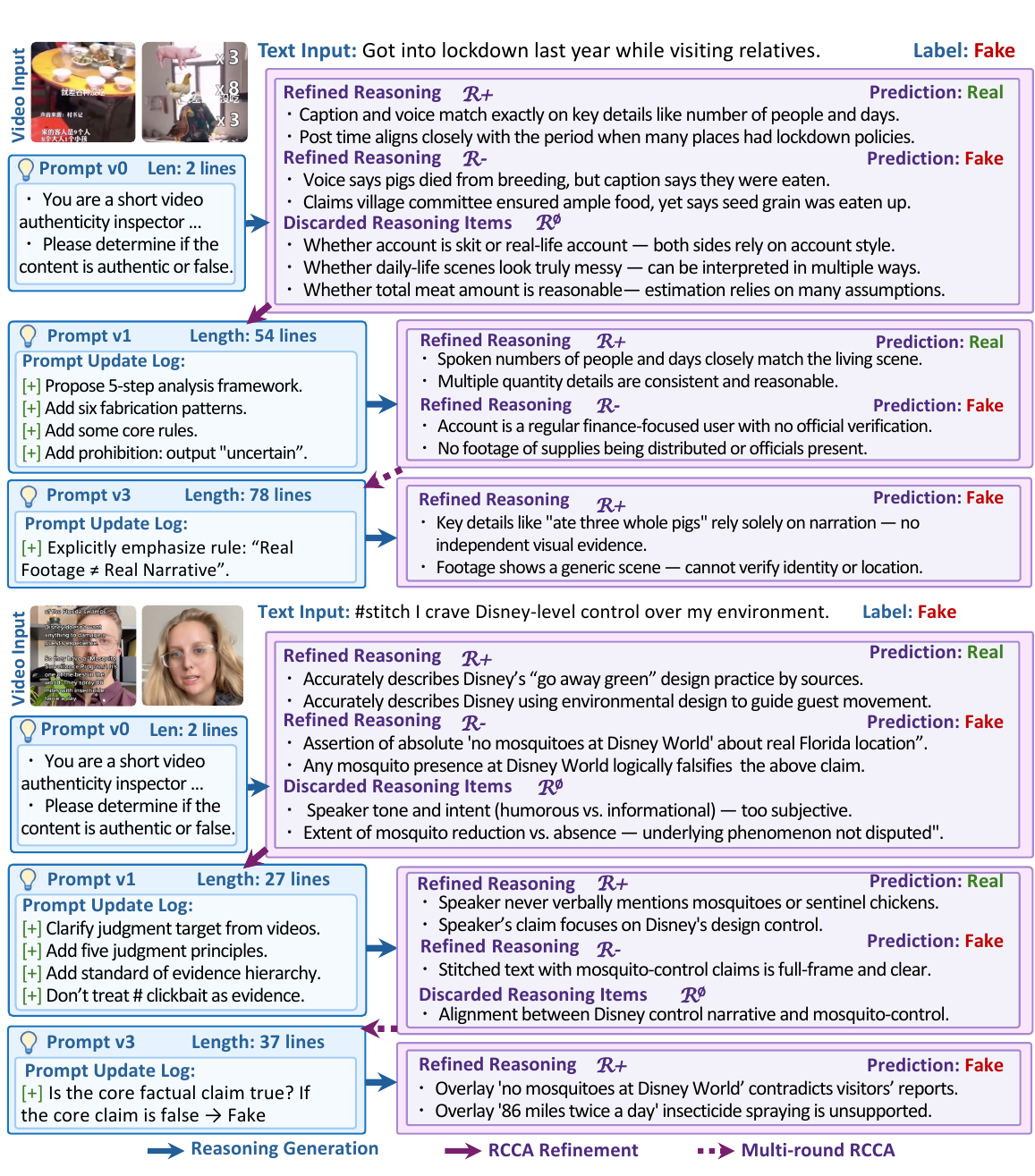}
  \caption{Case studies of the iterative prompt refinement process. For each example, we show the updated logs and the corresponding generated outputs after the initial iterations, as well as the key iteration at which the agent's prediction is corrected.}
  \Description{Two case studies show prompt updates, generated reasoning, and the iteration at which an initially incorrect prediction becomes correct.}
  \label{fig:case_study}
\end{figure*}

\noindent\textbf{Datasets and Baseline Settings.}
We conduct experiments on two widely used public short-video fake news
benchmarks, FakeSV~\cite{fakesv} and FakeTT~\cite{fakett_fakingrecipe}.
We follow the official chronological splits of both datasets, in which samples are
partitioned by video publication timestamp into training, validation, and test
sets with a 70\%/15\%/15\% ratio.
Table~\ref{tab:data_stats} shows that the two datasets have different topic
distributions but relatively balanced train--test splits within each topic.
Because the datasets contain historical events, cross-sample retrieval uses the
training set and does not involve online searches.
Additionally, to prevent the fine-tuned models from overfitting to a specific language, we perform
linguistic alignment during model preparation by translating the original
training materials from Chinese to English, and vice versa, including prompts, ASR results, and other textual inputs.

We primarily use the performance reported in the original baseline papers when
available.
Otherwise, we run their publicly available code following the implementation details in the corresponding papers.
For the raw VLM baselines, we apply our pre-iteration prompt to ensure a fair comparison.
Because some baselines are not open source and their implementation details are incomplete, we do not report their cross-dataset performance.

\subsection{Results and Comparisons}

\noindent\textbf{In-Domain Results.} Table~\ref{tab:main_results} reports the main comparison results under both in-domain and cross-dataset settings.
With Qwen3-VL as the backbone, SRM-FND achieves 91.33 overall accuracy on FakeSV and 92.31 on FakeTT, improving over the strongest prior baseline FakeSV-VLM by 1.11 and 3.01 points, respectively.
With InternVL3.5, SRM-FND achieves 90.04 on FakeSV and 91.30 on FakeTT, also outperforming all prior baselines by a clear margin.
Table~\ref{tab:topic_results} further reports the per-topic performance.
With Qwen3-VL, SRM-FND achieves accuracy above 88\% on all FakeSV topics and above 90\% on all four FakeTT topics.
InternVL3.5 shows a similar trend, with accuracy above 86\% on all FakeSV topics and above 89\% on all FakeTT topics, confirming that topic-adaptive routing captures heterogeneous verification cues across content domains.

\noindent\textbf{Cross-Dataset Results.} Under cross-dataset transfer, SRM-FND (Qwen3-VL) achieves 76.20 overall accuracy on FakeTT$\rightarrow$FakeSV and 72.24 on FakeSV$\rightarrow$FakeTT, which are 13.70 and 9.24 points higher than the strongest prior baselines, respectively.
SRM-FND (InternVL3.5) achieves 64.94 and 70.23 on the two transfer directions, also outperforming most prior baselines despite a larger gap from the Qwen3-VL variant.
We see that Hard News and Life \& Trivia remain relatively stable, while Specialized Knowledge and Social Safety exhibit larger degradation in certain transfer directions, indicating that these topics are more sensitive to domain shift and event-specific variation.
The transfer is also asymmetric across directions, suggesting that the two datasets differ in topic composition, event distribution, and platform-specific cues, while SRM-FND still preserves comparatively strong generalization.

\begin{figure}[!t]
    \centering
    \includegraphics[width=\columnwidth]{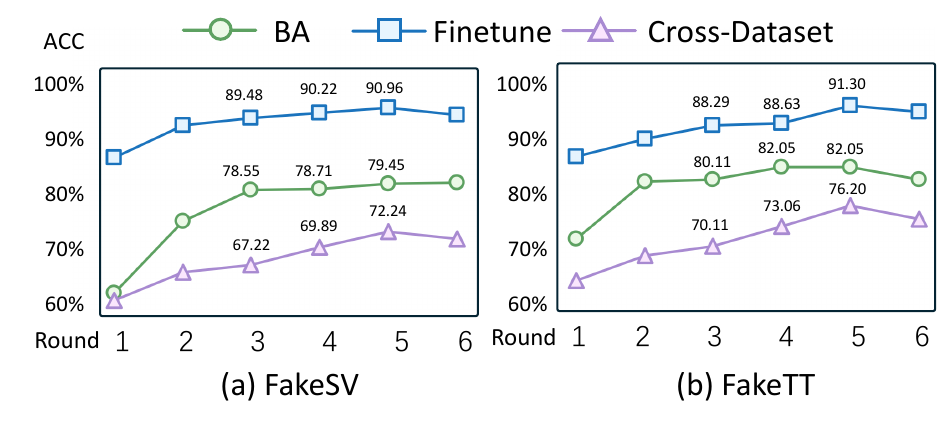}
    \caption{
    Accuracy curves of the Blind Analyst (BA) on the training set, the fine-tuned model on the in-domain test set, and the fine-tuned model on the cross-dataset test set.
   }
    \label{fig:accuracy_curve}
  \end{figure}

\begin{table}[t]
  \centering
  \caption{Ablation study of SRM-FND. ``Refine'' denotes contrastive CoT refinement, ``RCCA'' denotes iterative prompt refinement, ``Route'' denotes topic-adaptive expert routing, and ``Cross'' denotes cross-sample review.}
  \setlength{\tabcolsep}{1.1mm}
  \begin{tabular}{cccc|cc|cc}
    \toprule[1.5pt]
    \multicolumn{4}{c|}{\textbf{Settings}} & \multicolumn{2}{c}{\textbf{FakeSV}} & \multicolumn{2}{c}{\textbf{FakeTT}} \\
    \hline
    \textbf{Refine} & \textbf{RCCA} & \textbf{Route} & \textbf{Cross} & ACC & M-F1 & ACC & M-F1  \\
    \hline
     $\checkmark$ & & & & 83.39 & 82.97 & 83.61 & 82.27 \\
     $\checkmark$ & $\checkmark$ & & & 89.85 & 89.53 & 90.30 & 88.97 \\
      & & $\checkmark$ & & 90.22 & 89.94 & 90.64 & 89.32 \\
     $\checkmark$ & $\checkmark$ & $\checkmark$ & & 90.96 & 90.69 & 91.30 & 90.08 \\

    \hline
     $\checkmark$ & $\checkmark$ & $\checkmark$ & $\checkmark$ & \textbf{91.33} & \textbf{91.05} & \textbf{92.31} & \textbf{91.25} \\
    \bottomrule[1.5pt]
  \end{tabular}
  \label{tab:ablation}
\end{table}

\subsection{Analysis and Case Studies}

\noindent\textbf{Accuracy Trend over Iterations.} Fig.~\ref{fig:accuracy_curve} further shows the accuracy trend of the prompt-based Blind Analyst and the fine-tuned model across training and evaluation stages.
We observe that when the Blind Analyst reaches its peak accuracy, the corresponding fine-tuned model also attains peak in-domain and cross-dataset performance.
In particular, after about three iterations, the model already reaches good performance, and the gain from further iterations becomes increasingly marginal.
Nevertheless, later iterations still bring visible improvement in the cross-dataset setting, suggesting that the accumulated RCCA lessons gradually encourage more generalized reasoning than what is immediately reflected by in-domain samples.
This observation suggests that stronger Blind Analyst performance is associated with higher-quality generated CoT, because more samples can be correctly analyzed without relying on posterior labels as explicit knowledge, making such reasoning potentially more reliable than label-conditioned generation.


\noindent\textbf{Case Studies.} Moreover, we provide representative case studies in Fig.~\ref{fig:case_study} to illustrate two effects of the proposed method, namely, filtering double-edged evidence during contrastive deliberation and producing more label-aligned reasoning.
In the first case, the filtering step effectively establishes a clear contrast between supporting and refuting arguments while eliminating ambiguities.
In the second case, the presence of multiple speakers within the video initially confounds the model's reasoning and poses a challenge for RCCA to generate an appropriate prompt update.
Nevertheless, after two iterative refinements, the updated prompt successfully directs the agent to focus exclusively on the core claim, thereby resolving the ambiguity introduced by the multi-speaker context.

\noindent\textbf{Ablation Studies.}
Table~\ref{tab:ablation} reports the ablation results of SRM-FND.
First, we remove root-cause analysis and diagnostics from the refinement process, which leads to significant performance drops on both datasets, highlighting the importance of RCCA-guided refinement.
Second, we ablate topic-adaptive expert routing while retaining the full RCCA mechanism. In this setting, RCCA continues to enhance the base model through improved reasoning quality, but the absence of topic-aware routing limits the overall gains, indicating that contrastive CoT construction and RCCA refinement consistently benefit downstream detection.
Third, we remove both CoT refinement and cross-sample review, leaving only the adaptive routing mechanism, which yields only marginal improvements and suggests that routing alone is insufficient without high-quality reasoning.
Finally, we disable cross-sample review at inference time, resulting in a slight performance degradation, while the full model that integrates reasoning construction, topic-adaptive optimization, and cross-sample review achieves the best performance.
Even without cross-sample review, the model achieves state-of-the-art performance.

\section{Conclusions}

We propose SRM-FND, a self-reflective multimodal reasoning framework for short-video FND.
We build self-reflective, higher-quality CoT through contrastive deliberation with iterative root-cause diagnosis and corrective prompt refinement.
We further combine dual-phase topic-adaptive LoRA optimization with confidence-driven cross-sample review to improve both multimodal grounding and inference reliability.
Extensive experiments demonstrate the effectiveness, robustness, and interpretability of SRM-FND.

\begin{acks}
This work is supported by the China Postdoctoral Science Foundation under Grant 2025M771574,
the National Natural Science Foundation of China under Grant 62176260,
and the National Key Technology R\&D Program under Grant 2025ZD1700702.
\end{acks}



\bibliographystyle{ACM-Reference-Format}
\balance
\bibliography{sample-base}





\end{document}